# Towards an identity management solution on Arweave

Andreea Elena Drăgnoiu[a,b,c], Ruxandra F. Olimid[a,b]

[a]*Department of Computer Science, University of Bucharest, Romania, Bucharest, Romania*
[b]*Research Institute of the University of Bucharest (ICUB) , Bucharest, Romania*
[c]*certSIGN SA, Bucharest, Romania*

**Abstract**

Traditional identity management systems, often centralized, face challenges around privacy, data security, and user control, leaving users vulnerable to data breaches and misuse. This paper explores the potential of using the Arweave network to develop an identity management solution. By harnessing Arweave's permanent storage, our solution offers the users a Self-Sovereign Identity (SSI) framework that uses Decentralized Identifiers (DIDs) and Verifiable Credentials (VCs) to allow individuals and other entities to create, own, and manage their digital identities. Further, the solution integrates privacy-preserving technologies such as the BBS(+) signature scheme, thus enabling selective disclosure. This approach ultimately enhances user privacy and supports compliance with European Union legislation and regulatory standards such as the General Data Protection Regulation (GDPR) by design.

## 1. Introduction

A *digital identity* is a collection of electronically stored attributes and credentials uniquely identifying an individual, organization, or device in the digital world. These attributes can include personal information such as name, date of birth, social security number, biometric data, etc. Digital identities on blockchain assume decentralized systems for managing and verifying identities, which benefit from the transparency, security, and immutability of the

*Email addresses:* andreea-elena.panait@drd.unibuc.ro (Andreea Elena Drăgnoiu), ruxandra.olimid@fmi.unibuc.ro (Ruxandra F. Olimid)

underlying blockchain technology. Unlike traditional identity systems, which rely on centralized authorities (e.g., governments, corporations), blockchain-based digital identities can offer *self-sovereignty*, meaning that users have full control over their identity data.

*Motivation.* Arweave [1] emerged as a solution that offers a decentralized platform for the permanent storage of digital information, thus guaranteeing data availability for the future. The existing research on how Arweave network and protocols can be used in client applications is still in the incipient phase. Identity management (IdM) is necessary for many applications and a subject continuously open to research. Motivated by these facts, the necessity of introducing digital identity at a large scale, and the existing research on blockchain-based IdM solutions, we explore the possibilities of building an IdM solution on Arweave. Our proposal considers the regulations and standards concerning digital identities in the European Union (EU) [2, 3, 4, 5]. Thus, the current research is also motivated by the importance given to IdM nowadays, in the light of the EU Digital Identity Wallet [2], which is planned for full adoption across EU member states by 2026.

*Contribution.* To the best of our knowledge, we are the first to look into IdM solutions on Arweave. We propose a *Self-Sovereign Identity (SSI)* [6] management solution implemented on the Arweave test network using Verifiable Credentials (VCs) [7], Decentralized Identifiers [8], the Arweave Name System (ArNS) [9], and the BBS(+) signature scheme [10]. We exemplify the functionality by providing sample code in Javascript programming language. In terms of data disclosure, the security of the proposed solution reduces to the security of BBS(+). We compare the proposed solution with the most widely-spread (in terms of academic literature research and usage) IdM solutions, namely Sovrin [11], Veramo [12] (previously known as uPort), Civic [13], ION [14], and SelfKey [15]. We also discuss two variants that achieve minimal disclosure, in the sense of solely exposing the minimum of information necessary to pass identity verification (e.g., prove being major of age, i.e., age > 18 by EU regulations, but not expose the exact age). Finally, the presentation of Arweave and its functionalities is also a contribution, in the sense of an introductory tutorial for the general user.

*Outline.* The paper is organized as follows. Section 2 presents the related works. Section 3 gives the necessary background, including a quite detailed and technical presentation of digital identities, European regulations, and the necessary cryptographic primitives. Section 4 describes Arweave. Section 5 presents the overall design of the proposed solution and gives a possible implementation approach. Section 6 discusses the proposed solution in terms of properties, possible improvements, and comparison with other solutions. Section 7 concludes.

| Acronym | Explanation of acronym |
|---|---|
| ANS | Arweave Name Service |
| AO | Actor-Oriented |
| ArNS | Arweave Name System |
| BBS | Boneh-Boyen-Shacham |
| dApp | Decentralized Application |
| DNS | Domain Name System |
| DID | Decentralized IDentifiers |
| DLT | Distributed Ledger Technology |
| eID | Electronic IDentification |
| eIDAS | Electronic IDentification, Authentication, and trust Services |
| EORI | Economic Operator Registration and Identification |
| EU | European Union |
| EVM | Ethereum Virtual Machines |
| GRPD | General Data Protection Regulation |
| IdM | Identity Management |
| KYC | Know Your Customer |
| LEI | Legal Entity Identifier |
| MEM | Molecular Execution Machine |
| NFC | Near Field Communication |
| PoS | Proof-of-Stake |
| PoW | Proof-of-Work |
| SAML | eiDAS Security Assertion Markup Language |
| SEED | System for Exchange of Excise Data |
| SIC | Standard Industrial Classification |
| SSI | Self-Sovereign Identity |
| SPoRA | Succinct Proofs of Random Access |
| VC | Verifiable Credential |
| W3C | World Wide Web Consortium |
| zk | Zero-Knowledge |
| zk-proof | Zero-Knowledge Proof |
| zk-SNARK | Zero-Knowledge Succinct Non-Interactive Arguments of Knowledge |
| zk-STARK | Zero-Knowledge Scalable Transparent Arguments of Knowledge |
| zkVM | Zero-Knowledge-Virtual Machines |
| zkEVM | Zero-Knowledge Ethereum Virtual Machines |

Table 1: Acronyms

*Acronyms.* Table 1 lists the acronyms used throughout the paper.

## 2. Related Work

The existing work in IdM in general and blockchain-based IdM, in particular, is vast. Several surveys, including [16, 17, 18], present the recent state-of-the-art SSI blockchain-based development, whereas papers like [19] indicate various decentralized storage platforms available. We further restrict the discussion to solutions that are directly related to our current work or directly impact it (e.g., Arweave-related and similar blockchain-based IdM solutions used for comparison to our work).

The Ark Protocol [20] and Namespace.gg [21] are integral components of the Decent.land [22] ecosystem, each serving distinct roles in the management of decentralized identities and namespaces within the Arweave network [1]. The Ark Protocol [20] is a decentralized IdM framework, designed to facilitate the creation, management, and verification of digital identities, which integrates with the broader blockchain and Web3 ecosystems. It claims to enable secure, interoperable, and user-controlled identities. It provides standards and frameworks for DIDs, verifiable credentials, and interoperability [20]. Namespace.gg [21] is a decentralized naming service focused on the creation and management of unique namespaces. These namespaces function similarly to domain names in the traditional web but are designed for use in the decentralized Web3 environment. Namespace.gg can be combined with the Ark Protocol to create an identity system that integrates with Arweave: while Arweave provides the permanent storage backbone, Namespace.gg can be used for decentralized naming, and Ark Protocol can be used for managing multichain DIDs and credentials [20].

Sovrin [11, 16] is a global decentralized identity network built on Distributed Ledger Technology (DLT). It provides an infrastructure for issuing, storing, and verifying verifiable credentials. Sovrin uses open standards and a governance framework to ensure interoperability and trustworthiness across ecosystems. Its architecture emphasizes privacy by design, enabling selective disclosure and data minimization. As a public utility, Sovrin fosters a trusted digital identity ecosystem for individuals, businesses, and governments.

Veramo [12, 16] is a decentralized identity and credential management framework designed for building secure, self-sovereign identity solutions. It provides tools to create, manage, and verify digital credentials and DIDs using open standards such as W3C Verifiable Credentials and DID specifications. Veramo offers APIs and SDKs for developers to implement applications while ensuring user privacy and control.

Civic [13, 16] is a blockchain-based IdM platform that empowers users to control and protect their personal information. It enables secure, decentralized identity verification by allowing users to share only necessary data with trusted parties. Civic's app integrates biometric authentication and encrypted storage to ensure privacy and security. Businesses can leverage Civic's technology for streamlined Know Your Customer (KYC) and identity verification processes.

ION (Identity Overlay Network) [14] is a decentralized identity system built on Bitcoin, designed for managing DIDs. Developed by Microsoft, it operates as a Layer 2 solution, leveraging Bitcoin's blockchain for security while enabling scalable, fast DID operations without requiring new tokens or consensus mechanisms. ION allows users to create, update, and resolve DIDs, supporting self-sovereign identity principles. It uses open standards, including W3C-compliant DIDs and sidetree protocols, ensuring interoperability. ION promotes secure, decentralized IdM while maintaining compatibility with the Bitcoin ecosystem.

SelfKey [15, 16] is a blockchain-based IdM platform that provides tools for creating, verifying, and sharing identity documents with privacy and data minimization at its core. Through its SelfKey Wallet, users can store and manage identity-related documents and digital assets in a decentralized way. By emphasizing self-sovereignty, SelfKey empowers individuals and businesses to manage identity securely and efficiently.

## 3. Preliminaries

*3.1. European Regulations on Digital Identities*

In Europe, electronic identification is regulated by the Electronic IDentification, Authentication, and trust Services (eIDAS) [23], with the main goal of allowing electronic identification (eID) and secure access to online services within the EU and EEA countries [23]. eIDAS 2.0 Regulation [4] represents an update to the EU's 2014 eIDAS framework and provides additional rules for electronic identification and trust services to enable secure cross-border electronic transactions within the EU. The eiDAS Security Assertion Markup Language (SAML) Attribute Profile provides the list of attributes included in the eiDAS interoperability framework that supports cross-border identification and authentication processes [3].

For a natural person, the mandatory identity attributes required by the above-mentioned regulation are *Current Family Name*, *Current First Names*, *Date of Birth*, and *Unique Identifier*. Optional attributes include *First Names at Birth* or *Family Name at Birth*, *Place of Birth*, *Current Address*, *Gender*, *Nationality*, *Country of Birth*, *Town of Birth*, *Country of Residence*, *Phone no.*, *E-mail Address* [3].

For a legal person, the mandatory identity attributes required by the above-mentioned regulation are the *Current Legal Name* and the *Uniqueness Identifier*. The optional identity attributes include the *Current Address*, *VAT Registration No.*, *Tax Reference No.*, *Directive 2012/17/EU Identifier*, *Legal Entity Identifier (LEI)*, *Economic Operator Registration and Identification (EORI)*, *System for Exchange of Excise Data (SEED)*, *Standard Industrial Classification (SIC)*, *Legal Phone No.*, and *Legal E-mail Address* [3].

The technical specifications in line with the current regulations are given in terms of cryptographic requirements, SAML message format and attribute profile, and interoperability architecture [3].

Another EU law that establishes rules and guidelines for protecting individual data and privacy is the General Data Protection Regulation (GDPR) [5]. It sets strict rules on how organizations collect, store, and process personal data, emphasizing transparency, accountability, and individual rights. Its key principles include data minimization, accuracy, integrity, and purpose limitation. GDPR grants rights such as access, rectification, deletion, and portability of data to individuals. Organizations must gain consent, secure data, and report breaches within 72 hours. Non-compliance can lead to heavy fines of up to €20 million or 4% of global annual revenue. The regulation has influenced data protection practices worldwide, setting a global standard for privacy protection.

The EU Digital Identity Wallet [24] is a proposed digital identity solution by the EU that enables secure, verified, and convenient access to online services for EU citizens, residents, and businesses. It aims to provide a standardized, interoperable digital identity across all EU member states, supporting cross-border digital interactions. The wallet, accessible via mobile devices, allows users to securely store and share personal data. Core features include self-sovereign control, data minimization, and high privacy and secu-rity standards. It uses VCs, and DIDs, with options to leverage zk-proofs for selective disclosure of information [25]. The underlying design of the digital wallet is still under debate. An example in this respect is the cryptographers' feedback concerning the cryptographic methods used and the proposal to use mechanisms that were specifically designed to achieve the intended privacy and security goals [26]. By 2026, the EU countries must provide a digital identity wallet [2].

Last, but not least, the EU AI Act [27], expected to be finalized in 2024, would be the first of its kind to impact AI development and deployment. This regulation will have significant implications for digital identities, especially as digital identity systems increasingly incorporate AI-driven tools. Digital identity solutions, which utilize AI for biometric verification, risk assessment, and fraud detection, could be classified as high-risk AI under the Act due to their impact on fundamental rights and personal data security. Therefore, the Act aims to regulate AI applications that affect personal data and identity verification, ensuring these tools are used transparently and ethically.

### 3.2. Decentralized Identifiers and Verifiable Credentials

*Decentralized Identifiers.* A *Decentralized IDentifier (DID)* is a globally unique identifier that supports verifiable and decentralized digital IdM. A DID can refer to various entities such as individuals, organizations, objects, digital things, etc. The entity identified by the DID is called the *DID subject*, and together with other information associated with it (e.g., public keys, service endpoints, authentication protocols), is stored in the *DID document*. A *DID controller* is an entity that is authorized to make changes to the DID document. The DID controller might be specified in the optional *DID controller* property of the DID document, and its capabilities are defined in the *DID method*. Note that the controller might be the DID subject (but this is not always the case) and that a DID can have more than a single controller. The role of the DID controller, as a separate entity from the DID subject, is fundamental in the case of cryptographic key loss or key compromise [8].

The World Wide Web Consortium (W3C) provides a standardization for DIDs, called the *DID Core*, which includes the core architecture, data model, and representations [8]. The standard is universal so it can be applied to implementing DIDs on the Arweave network too. This assumes creating a DID document, storing it, and then resolving it by using the decentralized storage capabilities of Arweave and standard DID libraries [28].

*Verifiable Credentials.* Verifiable Credentials (VCs) are a digital method of proving claims about a person or an organization in a secure, privacy-preserving, and interoperable way [7]. A VC is formed of a *credential subject*, *claims*, *proof*, and *metadata*. Specific information (e.g., name, birth date) about the entity, called the *credential subject*, is a *claim*. The *proof* represents the cryptographic signature ensuring that the credential has not been tampered with. The details about the credential, such as the issuer, issuance date, and expiration date, represent the *metadata* of the VC. The entities involved in the VCs process are: the *issuer*, the *holder*, and the *verifier*. The *issuer* creates and signs the credential. The *holder* controls and stores the credential. The *verifier* validates the credential's authenticity. The W3C provides a standardization for VCs, called the *Verifiable Credentials Data Model* [7].

*3.3. BBS(+) Signature Scheme*

The BBS(+) [10] signature scheme, an improved version of the Boneh-Boyen-Shacham signature [29] (and the following BBS-type signatures [30, 31]), is a cryptographic scheme that can be used in VCs and identity systems. BBS(+) enables the signing of multiple messages using a single signature, resulting in smaller storage and faster verification times, which makes it efficient for the IdM use case. Moreover, BBS(+) accepts *selective disclosure*, i.e., the credential holder can prove ownership of a specific identity claim without revealing other claims. This allows for disclosing different identity claims to various parties, depending on the need. The BBS signature scheme [29] and its improvement BBS(+) [10] are proven to be secure under the well-established cryptographic assumptions Decisional Diffie-Hellman and q-Strong Diffie-Hellman [10]. The BBS(+) scheme is compatible with decentralized identity frameworks and can be integrated into systems that require revocation mechanisms [32].

Note that we do not need or intend to provide the mathematical description for the BBS(+) scheme here, as we further use it as a building block without the necessity to make its implementation details known to the reader. We further give a succinct explanation of how this cryptographic scheme is used. Figure 1 illustrates the main functionality of the BBS signature scheme family. We assume an issuer who signs a set of messages (step 1). The signature and the set of messages are securely sent to the holder (step 2). In our case, the holder is a user, and the messages are identity claims about the holder. The holder can generate proof about one or more of these messages with the following properties: the messages are disclosed, but no other information is revealed about the remaining undisclosed messages (step 3). The proof is called *zero-knowledge (zk)* because it reveals no information about the undisclosed messages. The scheme allows *selective disclosure* because the prover can select which messages to disclose. Finally, the proof and the disclosed messages are sent to a verifier (step 4), which checks for the validity of the proof (step 5). A valid proof attests that the disclosed messages were in the set initially signed by the issuer.

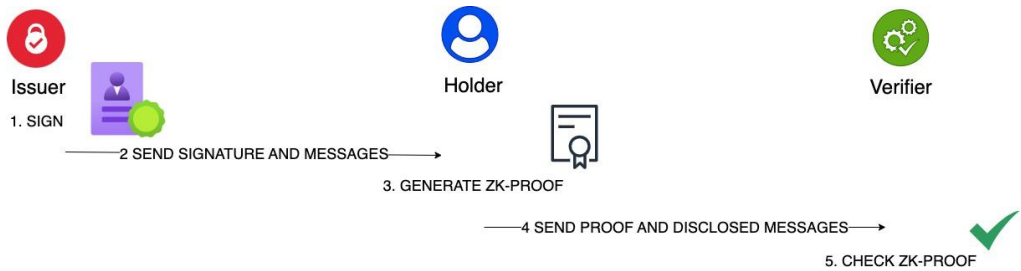

Figure 1: BBS(+) functionality overview (adapted from [33])

*3.4. Zero-Knowledge Proofs (zk-proofs)*

*Zero-knowledge proofs* (further referred to as zk-proofs) are cryptographic protocols that allow one party (called *the prover*) to demonstrate to another party (called the *verifier*) that he/she knows a certain piece of information (e.g., a secret, a solution to a problem) without revealing anything else about the private information up to a small, acceptable probability (normally, a negligible probability). Considering the communication between the prover and the verifier, there are two types of zk-proofs [34]:

1. *Interactive zk-proofs*: The prover and the verifier engage in a back-and-fourth communication process. The verifier challenges the prover, who responds with evidence that he/she knows the secret. Normally, both parties need to be online for the protocol to proceed.
2. *Non-Interactive zk-proofs*: The prover and the verifier do not have to interact. Instead, the prover generates a proof that can be independently verified by the verifier at a later time.

Lately, a particular type of non-interactive zk-proofs, called the *Zero-Knowledge Succinct Non-Interactive Arguments of Knowledge* (zk-SNARK) is widely used in blockchain implementations, because of its efficiency in terms of the proof size (the proof is succinct) and the verification time. A similar concept, the *Zero-Knowledge Scalable Transparent Arguments of Knowledge* (zk-STARK) presents some advantages (e.g., it does not require a trusted setup, it is more scalable than a zk-SNARK), but normally produces larger proofs, which make it less used in actual implementations [34].

*zk Implementations.* There are multiple zk tools and implementations, in particular for SNARKs and STARKs. Firstly, there are the pre-built reusable circuits, which are suited for developers with little zk expertise and smart contract applications that rely on usual zk building blocks, such as Merkle inclusion [35]. In this category, we mention Manta Network [36] and Semaphore [37]. Further, there are the zk Domain Specific Languages (DSLs), which can be used for minimal circuit sizes, and with little control over the proving backend. Examples include Circom [38], Aztec Noir [39], Cairo [40], ZoKrates [41], and Leo [42]. There are also libraries and frameworks for implementing zk-proofs, such as SnarkJS [43] and Halo2 [44]. Then, there are the low-level SNARK libraries, such as Arkworks-rs [45], that allow control over the entire prover stack and experimentation with different proving schemes, curves, or other low-level primitives. There are zk-compilers, such as Nil zkLLVM [46], useful to write highly customized circuits in a familiar language, but without the need to control the underlying cryptographic primitives [46]. Finally, there are the zk-Virtual Machines (*zkVMs*), such as the zk Ethereum Virtual Machines (*zkEVM*) which use succinct zk-proofs to prove the validity of an Ethereum state transition. Examples include ZKsync 2.0 [47], Polygon zkEVM [48], Scroll [49], Starknet [50]. Universal zkVMs also exist, for example, RiscZero [51] or zkWASM [52].

## 4. Arweave

In this section, we provide a basic description of the Arweave protocol and network so that the reader can understand the main functionalities. Note that the Arweave ecosystem is quite vast. Presenting all the notions related to Arweave is out of our scope. More information is available in the indicated references.

*4.1. Overview*

The *Arweave protocol* consists of a technical framework and set of rules that define how the Arweave system operates. This includes the specifications for data storage, retrieval, consensus, and the incentive mechanisms that power the network. Arweave enables permanent data storage, allowing users to preserve data with a one-time fee [28, 54].

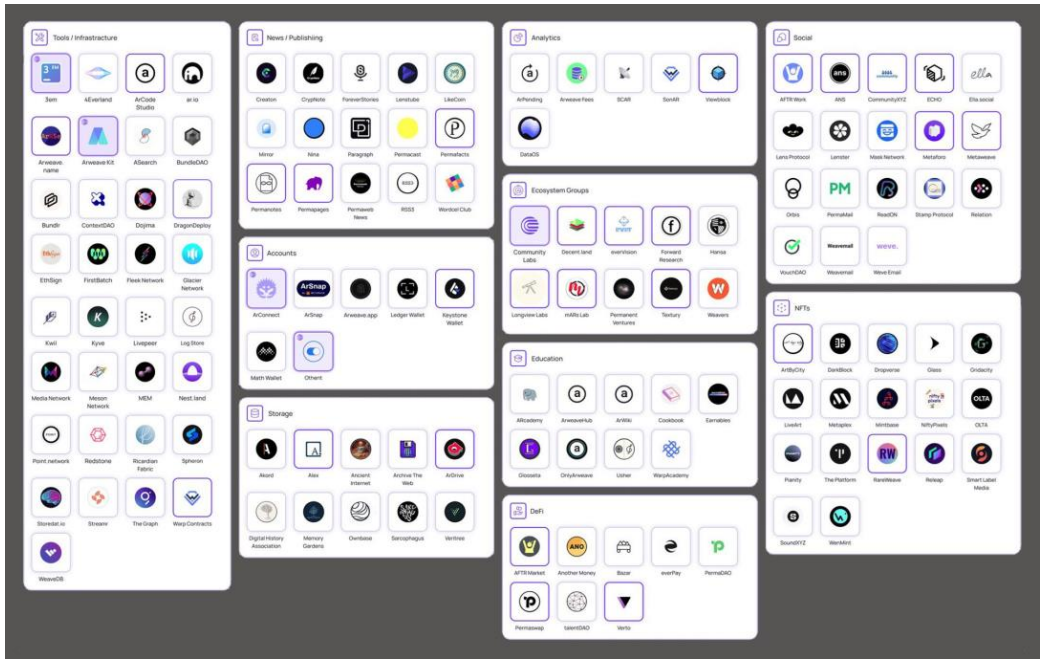

Figure 2: Arweave Eccosystem dApps as catalogued by the Community Labs [53]

The *Arweave network* is the decentralized network of computers (nodes) that implements the Arweave protocol. These nodes collectively store data, process transactions, and secure the network to create decentralized and permanent storage that allows the users to preserve data with a one-time fee. Permanent storage finds usability in many scenarios, including the persistence of art (e.g., literature, movies), facts (e.g., historical facts, opinions, reviews), and public ownership (e.g., NFCs). Arweave also builds *Permaweb*, a permanent and decentralized web inside the open Arweave ledger, including web pages, applications, etc. [1]. An overview of the most widely used decentralized applications built on the Arweave ecosystem, as cataloged by the Community Labs [53], is depicted in Figure 2.

*4.2. Mining and Consensus*

The protocol's execution involves a network of *miners* that are rewarded in AR tokens (the native token of the Arweave network) for storing data. The AR tokens incentivize the miners to store data, and by balancing CPU and storage resources to optimize mining performance, they increase their chances of being rewarded. Arweave's consensus is distinct from the traditional Proof-of-Work (PoW) or Proof-of-Stake (PoS), and it is facilitated through the use of *Succinct Proofs of Random Access (SPoRAs)*, which are critical to ensuring that data is widely distributed and stored by miners. The core idea behind Arweave's SPoRA consensus mechanism is that miners must demonstrate not only that they performed some computational hard work, but also that they have access to a random chunk of previously stored data in the Arweave network. This incentivizes miners to store and maintain access to historical data in the network [54, 55, 56].

*4.3. Endowment Model*

The Arweave endowment model consists of several components. First, is the *one-time payment model*, in which the users pay a one-time fee to store their data permanently. This fee is split between an immediate payment to miners and a contribution to a storage endowment. A second component is the *endowment fund*, a decentralized financial mechanism that generates interest over time. The interest earned is used to perpetually incentivize miners to store and maintain the data. For data storage, the overall Arweave network can be perceived as a permanent storage solution, whose maintenance involves miners who verify transactions, store data, and thus ensure that the solution remains secure and operational. The final component ensures the *economic sustainability* of the endowment model: as storage technology improves and the cost of storage decreases, it ensures that the incentive to store data remains economically viable. This creates a sustainable cycle where the initial payment and the accrued interest continuously support the permanence of the data [55, 54].

*4.4. Blockweave*

Arweave's underlying blockchain-like structure is called the *blockweave*. Blockweave is a data structure where not all nodes have to store the entire history of transactions. This is possible through memoization techniques and data structures that encapsulate all necessary information within individual blocks [1]. In a typical blockchain, each block is linked to the previous block, which leads to a linear sequence. As a difference, blockweave allows each block to reference multiple previous blocks, creating a *weaved* structure. In particular, a block is linked to its previous block, but also to (at least) one other older block in the chain. This enables better distribution and replication of data across the network. The structure ensures that miners must access and store data from older blocks to add new ones to the weave. This creates multiple copies of data, spread across different points in the network, ensuring that no single node holds all the responsibility for any one piece of data. By referencing more than one previous block, the blockweave ensures that data is redundantly stored and can be more efficiently accessed and verified [28].

*4.5. Permaweb*

The layer built on top of Arweave that allows for the creation and storage of web pages, applications, and other digital assets is called the *Permaweb*. These assets are accessible via unique transaction IDs or URLs. The Permaweb is formed of a set of protocols and consists of two fundamental components: (1) the blockweave that makes use of the SPoRA consensus mechanism (see Subsections 4.2 and 4.4) and (2) a sustainable endowment to ensure permanence of data, as previously explained (see Subsection 4.3) [55, 54]. The Permaweb is fully decentralized, allowing users to upload and store data. The data stored on the Permaweb is immutable: once uploaded, the data cannot be deleted or modified, providing a permanent record of websites, files, and applications. Permaweb applications and websites operate like normal web applications, except they are hosted in a decentralized, permanent environment. Content on the Permaweb is accessible using regular web browsers like Chrome or Firefox [54, 1].

*4.6. Smart Contracts*

Arweave supports several types of smart contracts, such as *SmartWeave* [57], *Warp Contracts* [58], *Molecular Execution Machine(MEM)* [59], or *WeaveWM smart contracts* [60]. We further briefly present each of them.

*4.6.1. SmartWeave*

*SmartWeave* [57] is the first smart contract protocol built on top of Arweave. It is designed to execute smart contracts off-chain and in a lazy evaluable approach. Unlike traditional blockchain smart contracts, for which every node in the network executes the contract code, SmartWeave introduces a new approach where contract execution is offloaded to the users interacting with the smart contract. SmartWeave contract's final state is reached by evaluating all state transitions (transactions) against the smart contract source code and the initial state. The execution of the contract logic off-chain reduces the on-chain computational overhead and associated costs, making it more scalable compared to the on-chain execution models. Hence, SmartWeave offers a straightforward model where the contract's initial state, contract source code, and state transitions (transactions) are stored on-chain, while execution runs off-chain, which can simplify development and reduce deployment costs. SmartWeave dApps often rely on centralized state indexers to offer better developer and user experiences [57].

*4.6.2. Warp Contracts*

*Warp Contracts* [58] offer enhanced development tools and a more developer-friendly interface compared to SmartWeave, facilitating easier contract creation and management. They allow for complex contract logic by handling pre- and post-execution operations, which can optimize and validate contract execution more effectively. Warp Contracts are designed to improve performance and scalability over SmartWeave by optimizing how contract logic and state are handled [58].

*4.6.3. Molecular Execution Machine (MEM)*

*Molecular Execution Machine* [59] provides a highly optimized execution environment, making it suitable for applications requiring intensive computational resources or high throughput. MEM supports a wide range of contract types and complex logic, offering greater flexibility for sophisticated applications. MEM implements performance enhancements and optimizations that can improve the efficiency and speed of contract execution and native smart contract state indexing in the cloud. As an evolving framework, MEMs might have limited adoption, tooling, and community support compared to more established solutions [59].

*4.6.4. WeaveWM*

*WeaveWM smart contracts* [61] operate in a custom virtual machine designed specifically for the Arweave blockchain. This environment allows for running more complex, data-intensive logic compared to traditional smart contracts on networks like Ethereum. A recent integration between WeaveVM and Arweave introduced the first-ever EVM precompiles designed to enable bidirectional data communication between the Ethereum Virtual Machine (EVM) and Arweave's permanent storage. In simple terms, these precompiles allow Solidity-based smart contracts on WeaveVM to both upload data to and read data from Arweave natively, bridging Ethereum's smart contract capabilities with Arweave's decentralized data storage. This solves the traditional EVM storage limitations, enabling developers to utilize Arweave's storage directly within the EVM environment [61].

| Protocol | Overview | Scope | Usability |
|---|---|---|---|
| ANS | Human-readable names for wallet or contract addresses | Simplify wallet and contract address interactions | Mostly used for general name-to-address resolution |
| ArNS | Framework for creating custom, application-specific naming systems | Offer flexible naming solutions for decentralized applications and platforms | More developer-oriented for specialized use cases |

Table 2: Comparison between ANS and ArNS

*4.7. Arweave Name Service (ANS)*

The *Arweave Name Service (ANS)* is a decentralized naming system built on the Molecular Execution Machine (MEM). It provides a way to associate human-readable names with Arweave wallet addresses or other resources on the Arweave network such as transaction IDs. This system simplifies interactions with the Arweave ecosystem by allowing users to reference wallet addresses or data using easily recognizable names instead of complex hashes or addresses. Thus, users can register and manage easy-to-remember names that map to Arweave addresses or content. ANS protocol uses the ".ar" as a top-level domain for its namespace domains. For example, "abc.ar" can map to a cryptographic wallet address, where wallet addresses are 43 characters long [62, 22].

*4.8. Arweave Name System (ArNS)*

The *AR.IO Protocol* [63] on Arweave is a decentralized infrastructure designed to enhance how users and developers interact with the Arweave blockchain by providing tools for more accessible, scalable, and user-friendly content storage and retrieval. The protocol uses a network of incentivized gateways that enable users to access data stored on the Arweave blockchain through standard web browsers using HTTP. This simplifies the interaction between users and the decentralized web by making Arweave's data available via URLs, such as https://arweave.net/txid, where txid is the transaction ID of the data. The *Arweave Name System (ArNS)* enabled by the AR.IO network of gateways, acts as a decentralized naming system similar to the traditional Domain Name System (DNS) to name Permaweb decentralized applications, pages, and data [9, 64]. ArNS is used to manage decentralized identifiers and namespaces directly on Arweave, with help from the Actor-Oriented (AO) supercomputer [65].

Table 2 presents ANS and ArNS in comparison, highlighting differences in scope and usability.

## 5. Solution Design and Implementation

We further discuss an IdM solution on Arweave in terms of design decisions, requirements, and implementation.

*5.1. Design Decisions*

We first discuss two design decisions related to privacy and security that should be taken into account when building an IdM solution. There are, of course, many other decisions to take in implementing such a solution, including the overall software architecture, election of tools and frameworks, etc., but we leave them out of the purpose of this paper.

*Identification Data.* The first step is to determine what identity data will be considered (government-issued IDs, biometrics, social security numbers, etc.) and how the data will be organized into identification fields and attributes. In our case, we consider the identity attributes in light of the eIDAS SAML Attribute Profile [3], as explained in Subsection 3.1.

*Data Protection.* The second step is to decide on the protection methods to be used. According to [8], to protect personal data, the DID Document associated with the digital identity should not contain sensitive encrypted data. If there is a need for encrypted data to be included, one should not associate any information that might be correlated with the identity.

In general, storing encrypted sensitive data on blockchain raises security concerns, and it is not recommended. This holds even more in the case of Arweave, for which storage is claimed to be permanent. This is because, in time, computational encryption weakens (encryption algorithms get broken, cryptographic keys' length becomes insufficient, etc.), and sensitive data gets exposed. In general, cryptographic mechanisms and parameters must comply with current recommendations, such as those mentioned in Subsection 3.1.

*5.2. Requirements*

After studying the eIDAS Regulation [23, 3, 4], General Data Protection Regulation (GDPR) [5], and the EU Digital Identity Wallet [24], we have identified some fundamental requirements for IdM. We further list these requirements, split into (1) *functional* requirements and (2) *privacy and security* requirements.

*Functional Requirements.* We have identified the following main functional requirements [4, 7, 8, 16, 23, 25, 66, 67]:

1. *Correctness*: the application should work as intended;
2. *Availability*: the solution has to be operational, accessible, and functional at need;
3. *User consent and control* (SSI) [4, 23, 16]: the users should manage (i.e., create, update, and delete) their digital identities independently of centralized authorities[1];
4. *Improved user experience* [4, 23, 16]: in terms of identity verification and identity-based access to services, simplified verification process and universal access to identity are desired;
5. *Interoperability*: the solution should be compatible with different standards, such as W3C [7, 8] and ISO [67].

*Privacy and Security Requirements.* We have identified the following main privacy and security requirements [4, 5, 25, 66, 68, 27]:

1. *Certified data* [4, 25]: data stored in the wallet should be certified by trusted authorities (for validity and reliability purposes);
2. *Confidentiality and privacy* [4, 5, 25]: compliance with data protection regulations, such as GDPR, must be assured;
3. *Selective disclosure* [4, 5]: the solution should only provide the necessary information for a given action (e.g., attest that the user is major of age, but do not disclose the exact age);
4. *Integrity* [66]: accountability and transparency regulated by different acts (e.g., Data Act [69] ) should be in place.
5. *Anonymization*: interactions, authorities, attributes, etc. should remain anonymous, certifying authorities and service providers should be kept mutually anonymous;

---

[1]Up to some limitations, as discussed in Subsection 6.3.

6. *Revocation*: mechanisms to revoke access to the wallet (e.g., if the device is compromised) should be set in place;
7. *Unlinkability* [68]: no one can trace or link how the digital identities are used, including coalitions of colluding parties (that should not be able to gather more information about a holder of a VC than what they individually already know);
8. *Cryptographic requirements*: the cryptographic keys should be securely stored, data should be transmitted through secured channels, cryptographic means should comply with up-to-date recommendations, etc.

*5.3. Actors*

In accordance with Section 3, we identify and fix the following actors:

1. *Issuer* : The entity that generates the VCs. The issuer normally represents a certified authority (e.g., governments, institutions).
2. *Holder* : The entity that owns some attributes and wants to obtain VCs. The holders are normally the users.
3. *Verifier* : The entity that verifies the authenticity of the holder's credentials. Based on this verification, the holder can, for example, authenticate in an application or gain access to a service. The verifier can be a (trusted) third party.
4. *Service provider:* The entity that provides the holder (user) with a service, if his/her identity is correctly verified by the verifier. The service provider and the verifier can be (and often are) the same entity.

Identity attributes are also defined as in Section 3 (e.g., name, date of birth, address).

*5.4. Implementation*

We now exemplify a possible implementation of IdM on Arweave, indicating all the necessary steps. We consider three phases: *Phase 1 - Setup and registration*, *Phase 2 - VCs and zk-proofs generation*, and *Phase 3 - Identity verification*. Phase 1 deals with registering the parties, including the generation of the keys and the creation of the DID documents. Phase 2 focuses on generating the VCs and the zk-proofs used to attest identity claims. Phase 3 consists of verifying the identity claims. For each phase, we provide sample code written in Javascript for exemplification. Figure 3 illustrates the overall process.

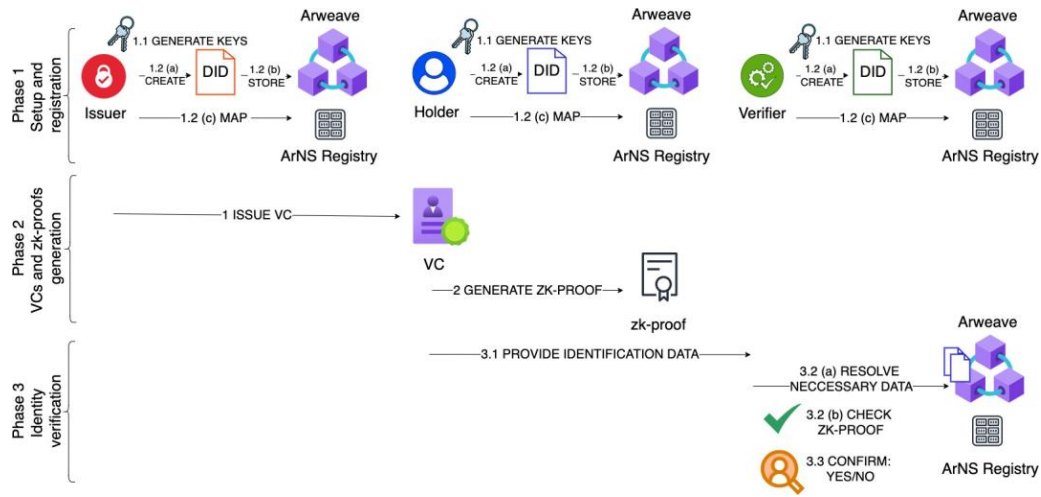

Figure 3: Overview of IdM framework on Arweave

*Phase 1: Setup and registration.* The first phase deals with the setup and registration of the parties.

1. *Generate key pairs.* Each entity generates a public-private key pair. An example of how key generation can be performed follows. For key management, one might use the ArConnect [70] or ArDrive [71] wallets.

```
const crypto = require('crypto');
const { BbsBls12381G2KeyPair,
        BbsBls12381G2Signature } =
      require('@digitalcredentials/bbs');
const { JwtSigner, JwtVerifier } =
      require('@digitalcredentials/did-jwt');

async function generateKeyPair() {
    const keyPair =
        await BbsBls12381G2KeyPair.generate();
    return keyPair;
}
```

2. *Issue and store DIDs.*
   (a) *Create DIDs.* Each entity creates a DID document containing the corresponding public key. An example of a DID document in JSON format follows.

```
const didDocument = {
  "@context": "https://www.w3.org/ns/did/v1",
  "id": "did:arweave:123456789abcdefghi",
  "verificationMethod": [
    {
      "id": "did:arweave:alice#keys-1",
      "type": "Ed25519VerificationKey2018",
      "controller": "did:arweave:123456789abcdefghi",
      "publicKeyBase58": "GfH1x23..."
    }
  ],
  "authentication": [
    "did:arweave:123456789abcdefghi#keys-1"
  ],
  "service": [
    {
      "id": "did:arweave:123456789abcdefghi#vcs",
      "type": "VerifiableCredentialService",
      "serviceEndpoint": "https://..."
    }
  ]
}
```

(b) *Store DID Documents on Arweave.* Each entity stores the DID document on Arweave, using *ArConnect* [70] or *ArDrive* [71]. To store the DID Document, one needs to create a transaction, add appropriate tags, sign the transaction using ArConnect or ArDrive, and post it to the Arweave network. The Arweave transaction ID can be further used as a reference to retrieve the document later. An example using the *arweave-js SDK* follows.

```
const Arweave = require('arweave');
const arweave = Arweave.init({
    host: 'arweave.net',
    port: 443,
    protocol: 'https'
  });
const storeDIDDocument = async (didDocument) => {
// Create a transaction to store the DID Document
const transaction = await arweave.createTransaction({
data: JSON.stringify(didDocument),
    });
// Add tags to identify the DID Document
transaction.addTag('Content-Type', 'application/json');
transaction.addTag('DID-Type', 'did-document');
// Sign the transaction using ArConnect
await window.arweaveWallet.signTransaction(transaction);
// Post the transaction to the Arweave network
const response
    = await arweave.transactions.post(transaction);
console.log('Transaction ID:', transaction.id);
return transaction.id;
```

```
        };
        // Store the DID Document
        storeDIDDocument(didDocument).then(txId => {
            console.log('Stored DID Document with
                transaction ID:', txId);
        });
```

(c) *Map ArNS names to Arweave transaction IDs*. Each entity registers a name within the ArNS Registry and maps the ArNS name to the Arweave transaction ID that corresponds to the DID document. For example, the holder registers the name alice and sets a pointer to the corresponding Arweave transaction ID (i.e., sets the pointer to the txID of the previously created DID document). Thus, assuming that the txID is *3t8YH9c2sN2F6GpOYXk...*, instead of using the unfriendly URL *https://arweave.net/3t8YH9c2sN2F6...*, the holder can simply use *https://alice.arweave.net*.

*Phase 2: VCs and zk-proofs generation.* The second phase is in charge of generating the VCs and the zk-proofs that attest identity claims.

1. *Issue VCs*. The issuer uses the BBS(+) signature scheme to sign the holder's credentials.
   (a) *Create credentials*. The issuer generates the credentials for the holder, which contain one or more attributes and their certification. The VC also contains the holder's public key. An example of how to generate a specific VC for Alice, aged 25, follows.

```
        const { generateBls12381G2KeyPair, sign, createProof, verifyProof } =
        require("@mattrglobal/bbs-signatures");

        async function createVerifiableCredential
        (issuerKeyPair, holderPublicKey,holderAttributes) {
            const credential =
            {
              "@context":
              ["https://www.w3.org/2018/credentials/v1"],
              id: 'https://example.org/credentials/123,
              type: "VerifiableCredential",
              issuer: "https://example.org/issuer",
              issuanceDate: new Date().toISOString(),
              credentialSubject: {
                id: "https://example.org/holder",
                "name": "Alice",
                "age": 25,
                ...,
                publicKey: holderPublicKey
                }
            };
```

(b) *Sign credentials.* The issuer signs the credential using the BBS(+) signature scheme. The signature will be embedded in the *proof* field of the VC, which is securely transmitted to the holder. An implementation example for the signature process follows.

```
const signedVC =
    await createVerifiableCredential(issuerKeyPair,
        holderKeyPair.publicKey, holderAttributes);
const signer = new JwtSigner({ key: issuerKeyPair });
const signedVC = await signer.sign(credential);
return signedVC;

credential.proof = {
  type: "BbsBlsSignature2020",
  created: new Date().toISOString(),
  proofPurpose: "assertionMethod",
  // this represents the issuer's DID key
  verificationMethod: "did:example:issuer#key-1",
  proofValue: signedVC.signature.toString('base64')
};
```

2. *Generate zk-proofs for specific claims.* The holder of the VC generates a zk-proof for a specific claim. A sample code exemplifying the disclosure of the age follows.

```
const { createProof } = require("@mattrglobal/bbs-signatures");
//Disclose only the age claim (index 2)
    const disclosedIndexes = [2];
    const zkProof = await createProof({
        signedVC,
        publicKey: holderKeyPair.publicKey,
        messages: credential.credentialSubject,
        revealed: disclosedIndexes
    });
```

*Phase 3: Identity verification.* The third phase deals with the verification of identity claims. First, the holder provides the verifier with all the necessary identification data. Second, the verifier checks the validity of the provided claims.

1. *Provide identification data.* To obtain a service or a product, the holder must first prove his/her identity data satisfies the necessary conditions. To accomplish this, the holder handles the verifier the ArNS name and the zk-proof computed on the VC data, in our example the proof of age.

2. *Verify identity data.*
   (a) The verifier retrieves the holder's DID document from Arweave via the ArNS, which resolves to the transaction ID corresponding to the DID Document.
   (b) The verifier further checks the zk-proof to validate the claim. A possible code for verification follows.

   ```
   const {verifyProof} = require("@mattrglobal/bbs-signatures");
   //Verify the proof
   const isProofValid = await verifyProof({
       zkProof,
       publicKey: keyPair.publicKey,
       messages: ["25"] // Disclosed claim
       });
   ```

   (c) The verifier confirms or not the user's identity claim based on the previous verifications.

For exemplification purposes and for simplicity of exposure, we have considered the age in our examples. Naturally, the date of birth is better suited, introducing more flexibility (the age changes in time) and being one of the mandatory identity attributes in the EU regulations (see Subsection 3.1).

## 6. Discussion

We further discuss how the proposed solution meets the identified requirements and additional benefits. Moreover, we give some implementation considerations that should be taken into account when implementing an IdM solution on Arweave and discuss the disclosure of more fine-grained, better privacy-preserving claims. Finally, we provide a brief comparison between our proposed solution and similar blockchain-based IdM solutions.

### 6.1. Solution Analysis

We discuss different properties in relation to our proposed solution. Some were already mentioned as requirements in Section 5, while others can be seen as additional benefits. We omit the strictly functional specifications, as these should be considered in precise settings and under a complete functional working scenario.

*Privacy and security.* The DID documents comply with the requirements in the sense that they contain public information only. The claims on sensitive attribute values (personal information such as name, date of birth, contact details, etc.) are correctly maintained in the VCs and certified by a trusted authority. This is twofold: first, the sensitive data storage aligns with privacy regulations, and second, the user cannot fake claims to help him/her acquire services or products for which he/she is not eligible. The BBS(+) scheme assures that the users discard the necessary claims only, while other sensitive data are not revealed and remain fully hidden due to the zk property of the BBS(+) proof. This means that BBS(+) enables *selective disclosure* concerning the claims in the VC. Cryptographic keys can be securely stored in ArConnect [70] or ArDrive [71] mobile application wallet. DIDs, DID documents, and VCs leverage strong cryptographic principles, thus enhancing security. The cryptographic schemes used, such as the BBS(+), are proven secure in strong enough adversarial models, thus they do not diminish the overall security of the system. In conclusion, by construction, the proposal asks the holder to disclose the necessary claims only, while assuring by the BBS(+) zk-proofs that they reveal no information about the undisclosed claims. Privacy of data and selective disclosure support normative such as the GDPR [5].

To conclude, the data security in our proposal, as described in Subsection 5.4, reduces to the security of the BBS(+) scheme. This holds directly from the construction, under the assumptions of a trusted VC issuer (that does not disclose sensitive information) and the existence of secure channels (for transmitting the VC from the issuer to the holder).[2]

BBS(+) also facilitates *unlinkability*, in the sense that users prevent cross-service tracking and increase anonymization because they avoid revealing unnecessary information. However, this is not sufficient overall, because the public key maintained in the DID uniquely identifies the user and permits tracking of his/her actions. In the absence of other safeguards, the immediate solution to overcome this is to transform the public key into a temporary identifier used only once, for example by refreshing the holder's keys after each verification. As such a solution brings in considerable complexity (creation of a new holder's DID document and issuing the corresponding VC for each verification), unlinkability vs. efficiency trade-off should be considered

---

[2]The DID document contains no sensitive data, so it cannot leak sensitive information.

by case.

*Self-sovereignty.* The solution enables the user to create and manage his/her own identity. The solution uses selective VCs that enable the user to choose which claims he/she wants to disclose and with whom he/she wishes to share the identity; therefore, the user is the sole owner of the identity. Users can access their data and ask for corrections (performed by updating the identity attributes) and deletion of their identities. Data update and identity removal are performed by a trusted, certified authority. This complies with EU regulations.

One can argue that the proposed solution is not fully SSI, as it relies on trusted authorities to certify the identity attributes. However, regulatory compliance somehow contradicts the decentralized nature of an SSI solution, and implementing a solution that does not comply with regulations (in our case, with the EU digital identity standards) is of no use. Therefore, we assume the existence of one or several certification authorities (even in the case of SSI), but keep the implication of such authorities minimal in the IdM process.

*Arweave-related properties.* As a direct consequence of being stored on Arweave, which is an immutable and permanent decentralized storage solution, the DID documents persist and cannot be altered or deleted. Both ArNS and DIDs promote decentralization, thus reducing reliance on central authorities. Also, integrating ArNS with DIDs facilitates interoperability with other decentralized identity systems and protocols. ArNS provides human-readable names, making the system more accessible: human-readable names are easier to remember and use than complex DIDs. The verification process for VCs is fast for both the credential holder and the verifier. VCs can be stored in digital wallets, facilitating accessibility and usability. All these aspects contribute to an increased user experience.

*6.2. Minimal Disclosure*

As presented in Section 5, our proposal discloses the exact value of the attribute(s) in the claim. For example, if a service or product is intended for a major of age, the holder provides certified proof of his/her age, including disclosing the exact age. Of course, minimal disclosure is not met, in the sense that the holder does not need to reveal the exact age to the verifier, but he/she only needs to prove that the age is higher than 18 (by EU regulations).

For privacy purposes, a solution enabling minimal disclosure is of interest. Therefore, we give two possible variants of the initial solution that achieve minimal disclosure.

*Variant 1.* In Phase 2 - step 1, the issuer creates the VC such that it contains claims that directly allow minimal disclosure. In our example, the VC contains a claim that attests that the holder is major of age (age > 18). The rest of the solution remains unchanged. The drawback of such a solution is that it reduces flexibility: the holder needs to rely on the issuer to generate the exact claims required. Although it might seem too restrictive at first sight, such a solution can be accepted if unlinkability is also expected (see the discussion in Subsection 6.1). In this case, if the issuer needs to anyway create a new VC at each refresh of the holder's keys, the introduced overload can be perceived as insignificant. In terms of privacy, such a solution increases privacy in relation to the verifier (the disclosure of the holder towards the verifier is minimal), but it might loosen anonymization in relation to the issuer (the issuer knows the exact claim the holder has to prove to the verifier, which discloses more about the products or services the holder wants to use the claims for, to the maximum of an exact identification).

*Variant 2.* To achieve both minimal disclosure for the verifier and minimal holder-service/products linkability for the issuer, the holder could create a zk-proof to attest the minimal disclosure claim based on a more general claim previously certified by the issuer in the VC. In our example, the holder creates a zk-proof to attest that the age (date of birth) is certified by the authority that issued the VC and that this age is at least 18. For such an implementation, one could follow the same line as in [34]. More precisely, the zk-proof in Phase 2 - step 2 is replaced by a zk-proof that attests to the verifier the minimal-disclosure claim and the correctness of the underlying attribute based on the VC, without disclosing the initial claim in the VC but certifying the attribute on which the claim is built. Such a solution has the drawback of a significantly increased complexity, in the sense that the holder and the verifier have to be involved in a two-party zk-proof protocol, including a setup phase with a fresh proving-verification key pair.

Further, we give more details on this implementation variant. The VC holder generates a zk-proof for the intended claim, using the corresponding attribute(s) from the VC. By using zk-proofs (in particular zk-SNARKs for efficiency reasons), the users can prove claims on one or more identity attributes (e.g., that the holder is major of age, that the nationality is within the EU) without revealing the actual data (e.g., the exact age, the exact nationality). At the same time, the zk-proof has to attest the certification of the underlying attribute, as present in the VC. The verifier is further appointed to verify the claims made by the user. For this, a one-time setup phase takes place, during which a proving-verification key pair is established. We assume the existence of a secure channel such that the verifier securely transmits the proving key to the holder. Later, the verifier will check the validity of the zk-proof created by the user using the verification key. This proposed variant differs from the initial solution in Subsection 5.4 (illustrated in Figure 3), mainly in Phase 2. Figure 6.2 illustrates these changes. Note that the zk-proof from BBS(+) is now replaced by a general (SNARK) zk-proof, with a different purpose, as previously explained. Hence, the overall security depends on the security of the underlying SNARK too. Furthermore, as in [34], the proving-verification procedure can be facilitated by the use of smart contracts published on the blockchain. To achieve this, the verifier publishes a smart contract that will verify the identity claims when accessed by the holder (on an input containing the generated zk-proof). Refer to Subsection 4.6 for more information on smart contracts in Arweave.

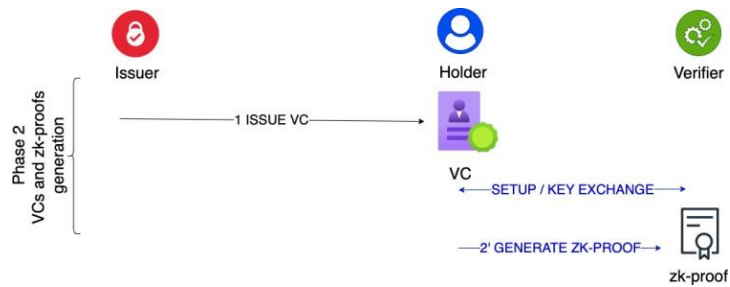

Figure 4: Variant 2 - Phase 2: Changes from the initial proposal (highlighted in blue)

### 6.3. Implementation Considerations

We further make some implementation considerations, as derived from the existing literature [8, 16, 54, 25, 72] and directly relate to our proposal.

*Privacy and security considerations.* As a general requirement, all implementations should comply with the up-to-date regulations and technical specifications. The solution should protect against potential attacks, such as name squatting or malicious DID document alterations. Moreover, ensuring the cryptographic security of the proofs (as well as of other cryptographic blocks) and the integrity of the Arweave data is vital.

To exemplify our proposal in Section 5, we have used the BBS(+) property to generate zk-proofs for user identity claims. For the first variant described in Subsection 6.2 there is no need for other cryptographic schemes, so no additional libraries or building-blocks implementations are required. For the second variant described in Subsection 6.2 there is a need for a specific zk-proof library. For efficiency reasons, zk-SNARKs are preferable. Our proposal references [34], which exemplifies with the ZoKrates [41]. However, ZoKrates might not be a suitable candidate for two reasons. First, it is a proof-of-concept implementation that has not been tested for production [41]. Second, the operations on the BLS12-381 elliptic curve (on Ethereum) are still under review, as suggested in [73]. The operations on the elliptic curve are important to efficiently perform BLS signature verification, which is necessary for certifying a user's claim. Performing on-chain calculations on Arweave is not natively supported, but it can be integrated using WeaveVM [60].

*Arweave-related considerations.* As a general requirement for assuring interoperability, the DID documents should follow the W3C DID specification. Moreover, the implementation must ensure that the mapping between ArNS entries and DIDs is consistent and securely maintained.

Wallet revoking access mechanism is a must. ArConnect offers the possibility to revoke access to a dApp or website by editing the offered permissions. ArDrive provides the option to encrypt files before uploading them on Arweave, and the decryption key can be kept secret. A limitation of these solutions would be that if someone has already saved or copied the decrypted content after accessing it, they will still have the content, and revoking the key only prevents further access by new users.

Storing data on Arweave incurs a cost, which is paid in AR tokens. One must ensure that users are aware of the transaction costs. User education is important for other reasons too. For example, educating users on the benefits and usage of DIDs and ArNS, as well as making the system user-friendly for non-technical users drives toward adoption. However, this is sometimes technologically limited (e.g., the computational cost of generating and verifying zk-proofs can be high).

Finally, scalability in terms of number of users and data transactions is important in decentralized IdM solutions. For dApps, scalability represents the ability of a blockchain network to handle an increasing number of transactions and users efficiently. On Arweave, this can be done by using bundled data. Bundled data (or simply *bundlers*) enable writing multiple independent data transactions into one top-level transaction [74]. Implementing such a solution means studying and implementing the ANS-104 standard [74]. In this way, IdM on Arweave could handle a large number of ArNS entries and DID resolutions efficiently. Therefore, we assume the existence of one or several certification authorities (even in the case of a SSI solution), but keep the implication of such authorities minimal in the IdM process.

| Solution | Blockchain | Privacy and Security | Scalability | Use Cases (Focus) |
|---|---|---|---|---|
| Our proposal | Public (Arweve) | DIDs, VCs, BBS(+) - selective disclosure, zk-proofs | High (with bundlers [74]) | IdM (general) |
| Sovrin [11] | Private | DIDs, zk-proofs | High | Government, financial services, healthcare |
| Veramo [12] | Public (Ethereum) | DIDs, VCs, selective disclosure | Limited by Ethereum | Voting, personal data management |
| Civic [13] | Public (Ethereum-based) | Biometric authentication, off-chain storage | Moderate | Personal/user IdM, financial services |
| ION [14] | Public (Bitcoin) | | High (layer-2) | Government, corporate IdM |
| SelfKey [15] | Public (Ethereum) | Off-chain data, encrypted pointers | Limited by Ethereum | KYC in financial and corporate sectors |

Table 3: Comparison with existing IdM blockchain solutions

*6.4. Comparison with Similar Solutions*

We briefly compare our solution with the most prominent blockchain-based IdM implementations: Sovrin [11], Veramo [12], Civic [13], ION [14], and SelfKey [15]. The comparison is conducted in terms of blockchain type (public vs. private), methods, and mechanisms to achieve privacy and security features, scalability, and primary use cases. Table 3 lists the findings. Most of the considered solutions are built on public blockchains, especially Ethereum. Some are focused on specific use cases, such as, for example,

SelfKey. From all, Veramo [12] (formally known as uPort Open, a solution we have previously analyzed in [72]) is similar in the sense that it also uses DIDs, VCs, and enables selective disclosure. More details on the considered solutions and a more in-depth comparison among them are available in the literature [16]. We did not identify any other IdM solution built on Arweave except our current proposal.

*6.5. Future Work*

The current paper gives a detailed description of the steps to be followed to build an IdM solution on Arweave. A more in-depth analysis of the security and privacy of such a solution, as well as implementation details and experimental analysis of a fully functional solution (also in comparison with the existing work, as a continuation of the brief comparison already provided), are the subject of our future work. For this, we aim for a strong collaboration with industry specialists who have experience with Arweave in the building of decentralized applications. In addition to the particular case of IdM, it is of interest to look at the applicability of Arweave for other use cases, test its limitations and gain a better understanding of its advantages.

## 7. Conclusions

We have provided an overview of the Arweave network, the protocols, and the applications that run on top of it. The existing academic research on Arweave is currently low. Hence, we consider our work significant for the readers interested in Arweave, but also the researchers and professionals interested in decentralized permanent storage solutions in general. Moreover, to our knowledge, this is the first paper to look at the (self-sovereign) identity management solutions over Arweave. The paper explored the development possibilities of the Arweave network by proposing an identity management solution, discussing implementation considerations, and highlighting its advantages and possible improvements. As decentralized applications evolve, together with the advances on the Arweave network, identity management solutions built on top of it can become more popular and beneficial among end-users and professionals within academia or industry.

*Acknowledgments.* This work was supported by a private scholarship offered by EntityC Consulting SRL to the first author. This work was supported by a grant of the Ministry of Research, Innovation and Digitaliza-tion, CNCS/CCCDI - UEFISCDI, project number ERANET-CHISTERA-IV-PATTERN, within PNCDI IV. The authors thank Mr. A.M.Voicu and his colleagues from EntityC Consulting SRL for their feedback and support concerning Arweave.